# Ferromagnetic Insulator to Metal Transition in Non-Centrosymmetric Graphene Nanoribbons

Aidan P. Delgado,[†,ǂ,‡] Michael C. Daugherty,[†,‡] Weichen Tang,[¶,‡] Steven G. Louie,[¶,§*] Felix R. Fischer[†,§,ǂ,f*]

[†] Department of Chemistry, University of California, Berkeley, CA 94720, USA.

[ǂ] Kavli Energy NanoSciences Institute at the University of California Berkeley and the Lawrence Berkeley National Laboratory, Berkeley, California 94720, USA.

[¶] Department of Physics, University of California, Berkeley, CA 94720, USA.

[§] Materials Sciences Division, Lawrence Berkeley National Laboratory, Berkeley, CA 94720, USA.

[f] Bakar Institute of Digital Materials for the Planet, Division of Computing, Data Science, and Society, University of California Berkeley; Berkeley, CA 94720, USA.

**ABSTRACT:** Engineering sublattice imbalance within the unit cell of bottom-up synthesized graphene nanoribbons (GNRs) represents a versatile tool for realizing custom-tailored quantum nanomaterials. The interaction between low-energy zero-modes (ZMs) not only contributes to frontier bands but can form the basis for magnetically ordered phases. Here, we present the bottom-up synthesis of a non-centrosymmetric GNR that places all ZMs on the majority sublattice sites. Scanning tunneling microscopy and spectroscopy reveal that strong electron-electron correlations, leading to the Stoner magnetic instability, drive the system into a ferromagnetically ordered insulating ground state featuring a sizeable band gap of $E_g \sim 1.2$ eV. At higher temperatures, a chemical transformation induces an insulator-to-metal transition that quenches the ferromagnetic order. Tight-binding (TB), density functional theory, and GW calculations corroborate our experimental observations. This work showcases how control over molecular symmetry, sublattice polarization, and ZM hybridization in bottom-up synthesized nanographenes can open a path to the exploration of many-body physics in rationally designed quantum materials.

## INTRODUCTION

The engineering of low-energy modes in bottom-up synthesized graphene nanoribbons (GNRs) has become a versatile testbed for the realization and exploration of designer Hamiltonians in one-dimensional (1D) carbon nanomaterials.[1-6] While lateral quantum confinement in GNRs typically leads to the opening of a sizeable band gap,[7-10] the effective interaction of singly occupied, in-gap, localized electronic states—such as symmetry-protected topological interface states aligned along a GNR backbone—has been used to construct periodic lattices (i.e., bands) of zero-energy modes (ZMs).[4, 11-14] These half-filled bands can serve as the basis for engineering complex electronic and magnetic interactions in 1D carbon nanomaterials that can be described by the Su-Schrieffer-Heeger,[15, 16] Hubbard,[17] or spin Heisenberg models.[18] A careful tuning of hopping parameters has been shown to give access to narrow-gap insulators,[11, 12, 19, 20] intrinsically metallic GNRs,[13, 21] and magnetically ordered 1D spin chains.[22-26] Here, we draw inspiration from the Stoner model for itinerant electrons[27] (Figure 1A) to engineer a Hamiltonian that describes a Stoner metal-insulator phase transition in a non-centrosymmetric GNR. We demonstrate how the ratio between the kinetic energy, represented by the hopping integral ($t$), and the on-site electron-electron repulsion ($U$) can be controlled by molecular design. Control over the density of states at the Fermi level ($E_F$) provides access to both long-range ferromagnetically ordered insulating phases and metallic states within a single GNR architecture. The control over spin degrees of freedom in bottom-up fabricated GNRs represents a longstanding challenge in carbon-based nanomaterial technology and opens pathways toward GNR-based spintronic and quantum devices.

Our strategy for engineering both magnetically ordered insulating and paramagnetic metallic phases in a GNR builds on the criterion for the Stoner magnetic instability.[27] Within this framework, the ground state of a 1D system (of finite length) is determined by the product of $U$ and the density of states at $E_F$ ($D(E_F)$).

$$U \times D(E_F) > 1 \tag{1}$$

If this inequality is satisfied, the paramagnetic ground state is unstable, and the system can lower its total energy by developing a finite magnetization. In a model for a 1D chain, including only the nearest neighbor (NN) hopping (assuming $t > 0$ in the following discussion, which can always be accomplished in 1D using gauge transformation), the $D(E_F)$ is inversely proportional to the hopping amplitude.

$$D(E_F) = (2\pi t)^{-1} \tag{2}$$

The Stoner criterion for a magnetic transition can thus be expressed in terms of the ratio of on-site interaction and NN hopping kinetic energies:

$$U / t > 2\pi \tag{3}$$

Whenever $t$ is small, a large $D(E_F)$ induces the Stoner instability, favoring a spin-polarized (magnetic) insulating ground state (Figure 1B). Conversely, if $t$ is large enough so that $U / t < 2\pi$, the kinetic energy suppresses spin polarization, stabilizing a metallic ground state.

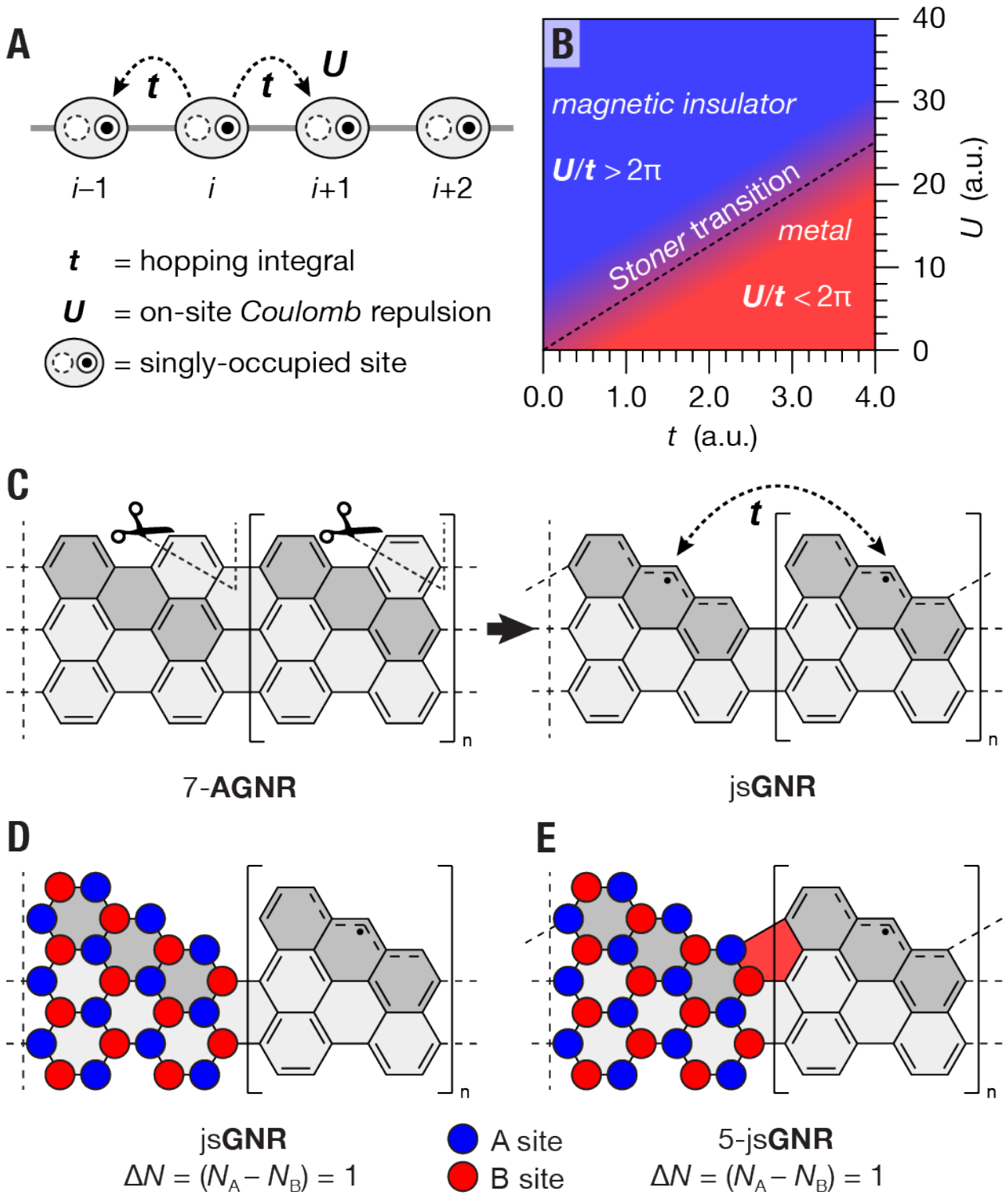

**Figure 1.** Engineering electronic and magnetic structures in non-centrosymmetric GNRs. (A) Hubbard TB model for a 1D chain at half-filling with on-site Coulomb repulsion ($U$) and symmetric hopping integral ($t$). Each hopping site (gray oval) corresponding to a localized zero-mode molecular state can host two electrons with opposite spin. (B) Phase diagram of the Stoner model for itinerant magnetism in a 1D chain ($U$ is given in the same arbitrary units of the symmetric hopping integral $t$). Metallic (insulating) phases are predicted for $U/t < 2\pi$ ($U/t > 2\pi$). (C) Schematic representation of the design of jsGNRs derived from the structure of 7-AGNR. Removal of three C-atoms per unit cell along one edge gives rise to a non-centrosymmetric ribbon lined by a pristine armchair edge on one side and a sawtooth zigzag edge on the other. (D) Lattice imbalance in jsGNRs gives rise to half-filled zero modes localized on the majority A sublattice. (E) Fusion of [4]helicene fragments in jsGNRs (red 5-membered ring) lining the sawtooth edges gives rise to sublattice mixing in 5-jsGNRs.

A functional translation of this model into the real space structure of a GNR is represented in Figure 1C. Starting from the expanded unit cell of a 7-AGNR (two anthracene cores per unit cell), we subsequently remove three C-atoms (an allyl group) per supercell from the same edge along the GNR backbone. The resulting dangling bonds are terminated by H-atoms, giving rise to the characteristic pattern of laterally fused [4]helicene fragments lining the top edge of Janus-type sawtooth GNRs (jsGNRs). The bottom armchair edge remains untouched. The short zigzag segments lining one side of jsGNRs not only break inversion and mirror symmetry but also introduce a sublattice imbalance (Figure 1D, $\Delta N = (N_A - N_B) = 1$, where $N_A$ and $N_B$ represent the number of atoms per unit cell residing on the intrinsic A and B sublattice sites of graphene, respectively). Ovchinnikov's rule[28] and Lieb's theorem[29] predict a single unpaired electron —and thus one ZM—per unit cell residing on the majority sublattice A, mirroring the TB model for a 1D Hubbard chain at half-filling. In the case of jsGNRs (Figure 1D), the effective hopping ($t_{eff}$) between these sublattice polarized ZMs is only mediated by second-nearest neighbor hopping (resulting in a small $t_{eff}$ and a large $D(E_F)$), favoring a correlated magnetically ordered insulating ground state if Eq. (3) is satisfied. An enhanced hopping regime in jsGNRs can be accessed by breaking the underlying bipartite lattice symmetry of graphene, e.g. by fusing [4]helicene fragments along the edges of jsGNRs into 5-membered rings represented by 5-jsGNRs in Figure 1E.[13] In the sublattice mixed case, contributions from nearest neighbor hopping increases $t_{eff}$ ($D(E_F)$ decreases correspondingly), stabilizing a metallic band structure. Careful tuning of $t_{eff}$ through chemical design thus allows the same molecular platform to host either metallic or insulating ground states.

## RESULTS/DISCUSSION

**Molecular Precursors and Surface-Assisted Growth of jsGNRs.** Guided by this idea, we designed the molecular precursor for jsGNRs, 9-iodo-10-(4-iodo-2-methylnaphthalen-1-yl)anthracene (**1**), depicted in Figure 2A (for molecular synthesis and characterization, see Supporting Information Scheme S1). Samples of jsGNRs were prepared following established surface-assisted bottom-up GNR synthesis protocols. **1** was sublimed in ultrahigh vacuum (UHV) from a Knudsen cell evaporator onto a clean Au(111) substrate held at $T = 25$ °C. Figure 2B shows a representative topographic STM image of a densely packed monolayer of **1** self-assembled into linear arrays. The step-growth polymerization of **1** was induced by annealing the molecule-decorated surface to $T = 180$ °C for $t = 15$ min. A second annealing step at $T = 300$ °C ($t = 15$ min) induced thermal cyclodehydrogenation, giving rise to the fused GNR backbone. Topographic STM images show the characteristic signatures of jsGNRs featuring two clearly differentiated edge structures (Figure 2C). A smooth GNR edge associated with the conventional armchair termination is contrasted by a periodic $z$-height modulation ($\Delta z \sim 69$ pm) on the opposing sawtooth edge (Supporting Information Figure S1). The superlattice of protrusions is commensurate with the size of the jsGNR unit cell, suggesting its origin can be attributed to a preferred adsorption geometry of [4]helicene fragments lining the sawtooth edge. While constant-height bond-resolved STM (BRSTM) imaging clearly resolves the pattern of 6-membered rings lining the armchair edge of jsGNRs (Figure 2E), the non-planar conformation adopted by the sawtooth termination precludes distortion-free imaging (this effect has previously been observed in metallic sGNRs).[13] Statistical analysis of jsGNRs samples shows no significant preference for the head-to-tail (HT) over the head-to-head (HH) alignment of monomer **1** during the step-growth polymerization, limiting the size of pristine HT-*poly*-**1** segments to < 10 repeat units.[30] The alternative tail-to-tail (TT) fusion resulting from the coupling of two anthracenyl groups during the polymerization was not observed (Supporting Information Figure S2).

A third annealing step at $T = 350$ °C ($t = 15$ min) induced the fusion of [4]helicene fragments lining the sawtooth edge (Figure 2A). The introduction of 5-membered rings breaks bipartite lattice symmetry (e.g., A to A site hopping is now allowed) and results in an effective coupling of the theoretically predicted spin-polarized A and B sublattice sites of graphene for the ZM states. An asymmetric ring-strain along the edges of 5-jsGNRs is reflected in the characteristic curved backbone (Figure 2D). Topographic STM images reveal a uniform $z$-height along either edge of the ribbon (Supporting Information Figure S1). Constant-height BRSTM images clearly resolve the 6-membered rings lining the convex armchair edge and a periodic pattern of 5- and 6-membered rings along the concave sawtooth edge of 5-jsGNRs (Figure 2E). The careful design of molecular precursors, along with the precise control of growth parameters throughout the annealing step, give access to a class of Janus-type GNRs whose electronic and magnetic properties are defined by a non-centrosymmetric crystallographic unit cell.

**Electronic Structure Characterization of jsGNRs and 5-jsGNRs.** Having established the selective on-surface growth of jsGNRs and 5-jsGNRs, we set out to probe their respective electronic structures using scanning tunneling spectroscopy (STS). Figure 3A shows representative d*I*/d*V* point spectra acquired above the position of the armchair edge (red) and the sawtooth edge (blue, orange) of a jsGNR (d*I*/d*V* spectra reflect the local density of states, or LDOS, under the STM tip). Differential conductance spectra recorded along the sawtooth edge show two prominent features bracketing $E_F$. A steep rise in the d*I*/d*V* signal at $V_s$ = +0.80 V (*feature* 1) is mirrored by a peak at negative bias $V_s$ = −0.50 V (*feature* 2), corresponding to the lower (upper) edge of an unoccupied (occupied) electronic band, respectively. At more negative bias, we observed a sequence of three minor features followed by a more dominant larger peak at $V_s$ = −1.60 V (*feature* 3). From the position of *features* 1 and 2, we can extract an electronic bandgap of $E_{g,exp}$ = 1.2 ± 0.1 V for jsGNRs on Au(111). This interpretation is further corroborated by differential conductance maps recorded along the backbone of jsGNRs. Across the range of −0.25 V < $V_s$ < +0.70 V within the gap, we only observed a diffuse featureless LDOS typical for semiconductors (Supporting Information Figure S3). d*I*/d*V* maps recorded at sample biases corresponding to *features* 1, 2, and 3 instead showed discrete nodal patterns (Figures 3B–D). Differential conductance maps recorded at $V_s$ = +0.80 V (*feature* 1, Figure 3B) and $V_s$ = −0.50 V (*feature* 2, Figure 3C) feature three bright lobes lining short zigzag segments along the sawtooth edge. The armchair edge remains featureless against the signal background. d*I*/d*V* maps recorded at biases corresponding to *feature* 3 ($V_s$ = −1.60 V, Figure 3D) show the emergence of a fourth lobe along the sawtooth edge accompanied by a broadening of the signal intensity.

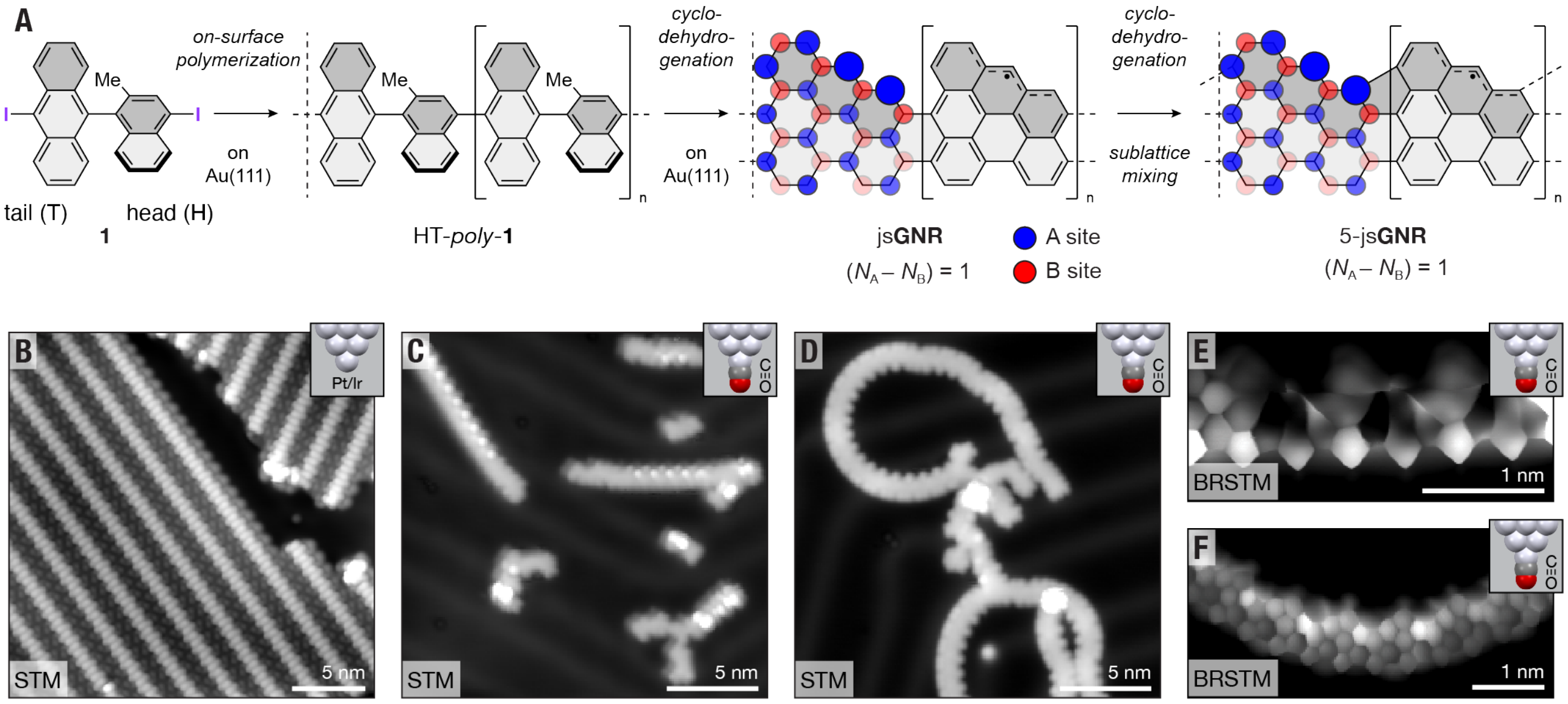

**Figure 2.** Schematic representation of the synthesis and on-surface characterization of jsGNRs. (A) Schematic representation of the stepwise on-surface synthesis of jsGNRs from **1** on Au(111). Thermally induced cyclodehydrogenation of jsGNRs gives rise to 5-membered rings in sublattice mixed 5-jsGNRs. (B) STM topographic image of self-assembled island of molecular precursor **1** thermally evaporated onto Au(111) (sample bias $V_s$ = −150 mV, tunnel current $I_t$ = 50 pA, CO-functionalized tip). (C) STM topographic image of jsGNRs on Au(111) after stepwise annealing at $T$ = 180 °C for 15 min and $T$ = 300 °C for 15 min. ($V_s$ = −800 mV, $I_t$ = 50 pA, CO-functionalized tip). (D) STM topographic image of 5-jsGNRs on Au(111) after annealing at $T$ = 350 °C for 15 min. ($V_s$ = −800 mV, $I_t$ = 20 pA). (E) Constant-height bond-resolved STM (BRSTM) image of a segment of jsGNRs following the initial cyclodehydrogenation sequence at $T$ = 180 °C and $T$ = 300 °C ($V_s$ = 5 mV, CO-functionalized tip). (F) Constant-height bond-resolved STM (BRSTM) image of a segment of 5-jsGNRs following additional annealing at $T$ = 350 °C ($V_s$ = 20 mV, $I_t$ = 150 pA, CO-functionalized tip). All STM experiments were performed at $T$ = 4.5 K.

To corroborate the assignment of spectroscopic features and gain further insight into the electronic structure of jsGNRs, we performed *ab initio* density functional theory (DFT) simulations for a freestanding jsGNR. To obtain a more accurate estimate of the band gap, we further carried out many-electron ab initio GW calculations to account for quasiparticle self-energy effects. The spin-resolved DOS and the corresponding band structure are shown in Figures 3H,I. The most prominent features in the DOS are a pair of peaks at $E$–$E_F$ = +0.9 eV and $E$–$E_F$ = −0.9 eV bracketing $E_F$. The exchange field induced by the sublattice and spin polarization of the zigzag edge state lifts the spin degeneracy and splits the zero-mode band (ZMB) into two spin-polarized subbands: the unoccupied zero-mode band (UZMB) and the occupied zero-mode band (OZMB). This sizeable splitting gives rise to a theoretically predicted band gap of $E_{g,GW}$ = 1.8 eV (as observed in previous studies, the experimental band gaps derived from STS tend to be smaller than theoretical predictions since the GW quasiparticle calculations for the freestanding structure do not include the additional dielectric screening from the underlying Au(111) substrate). At half filling, only the OZMB is occupied by electrons with their aligned spins contributing to the ferromagnetically ordered ground state of the jsGNRs. While theory predicts two narrow peaks in d*I*/d*V* spectra corresponding to the UZMB and OZMB, the large phase space for scattering and the exceptionally short lifetimes of quasiparticles injected into strongly correlated, nearly dispersionless flat bands leads to significant

Lorentzian broadening reflected in steps in the d$I$/d$V$ signal associated with *feature* 1 and 2.[26]

Figures 3E,F show the calculated LDOS maps at energies corresponding to the UZMB and OZMB. The simulated nodal patterns for the two bands are indistinguishable (they are composed of states with the same real-space orbital wavefunction and differ only in the spin degree of freedom) and in good agreement with the experimental d$I$/d$V$ maps supporting our assignment of spectroscopic *features* 1 and 2 to the UZMB and OZMB, respectively. Projection of the LDOS at $E$–$E_F$ = –1.8 eV (Figure 3G) supports the assignment of *feature* 3 to be the top of the next fully occupied band below OZMB. The correspondence between theory and experiment supports our hypothesis that the ground state of jsGNRs is a ferromagnetically ordered insulator.

To evaluate the impact on the electronic structure associated with the fusion of [4]helicene fragments into 5-membered rings along the sawtooth edges of 5-jsGNRs, we collected d$I$/d$V$ point spectra on a finite curved segment containing 16 unit cells (Figure 3J). Differential conductance spectra acquired above the position of the armchair edge (red) or the sawtooth edge (blue, orange) consistently show four prominent features: a sharp step in the d$I$/d$V$ signal at $V_s$ = +1.50 V (*feature* 4), two smaller peaks at $V_s$ = +0.15 V (*feature* 5) and $V_s$ = –0.70 V (*feature* 6) bracketing $E_F$, followed by a shoulder and gradual rise in signal intensity below $V_s$ = –1.50 V (*feature* 7). In stark contrast to the nonplanar jsGNRs, the d$I$/d$V$ signal in 5-jsGNRs does not fully decay to the background across the bias range of –0.70 V < $V_s$ < +0.15 V, suggesting a finite DOS at the $E_F$. Differential conductance maps recorded at biases spanning *features* 5 and 6 show complex nodal patterns across the entire width of the ribbon, evolving from bright dots at the position of 5-membered rings surrounded by chevron shapes (Figure 3K) to linear corrugated features lining the armchair edge (Figure 3M, Supporting Information Figure S4). Even at $V_s$ = 0.0 V (Figure 3L), prominent nodal structures reminiscent of an interpolation between *features* 5 and 6 could be observed. The persistence and gradual transition of nodal patterns spanning the full range of –0.70 V < $V_s$ < +0.15 V suggest that the electronic structure of 5-jsGNRs features a metallic band (with bandwidth $\Delta E_{exp}$ = 0.8 ± 0.1 V) crossing the $E_F$.

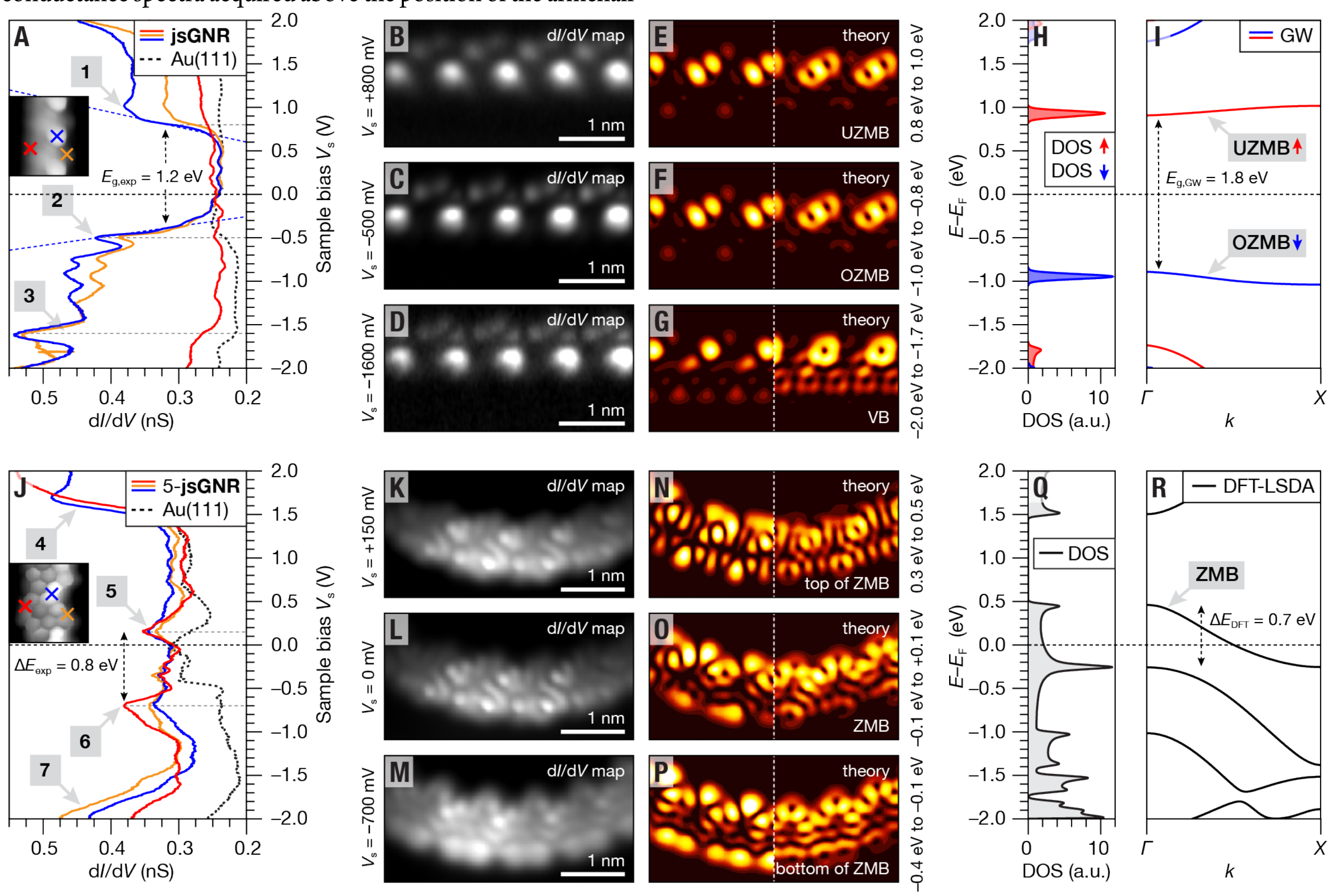

**Figure 3.** Electronic structure of jsGNRs and 5-jsGNRs. (A) d$I$/d$V$ point spectroscopy of jsGNRs on Au(111) recorded at the positions highlighted in the inset (spectroscopy: $V_{ac}$ = 10 mV, $f$ = 455 Hz; imaging: $V_s$ = 100 mV, $I_t$ = 20 pA, CO-functionalized tip). (B–D) Constant-height d$I$/d$V$ maps recorded at the sample voltage biases of $V_s$ = +800 mV, $V_s$ = –500 mV, and $V_s$ = –1600 mV, respectively ($V_{ac}$ = 10 mV, $f$ = 455 Hz, CO-functionalized tip). (E–G) DFT LDOS maps above a freestanding jsGNR at energies corresponding to UZMB, OZMB, and the top of the VB calculated using the model for a pure *s*-wave (left) and pure *p*-wave (right) STM tip (experimental d$I$/d$V$ maps acquired with CO-functionalized tips feature contributions of both *s*- and *p*-wave character). (H) Spin-polarized LDA-GW DOS for a freestanding jsGNR (spectrum broadened by 27 meV Gaussian). (I) Spin-polarized GW band structure of freestanding jsGNR. (J) d$I$/d$V$ point spectroscopy of 5-jsGNRs on Au(111) recorded at the positions highlighted in the inset (spectroscopy: $V_{ac}$ = 10 mV, $f$ = 455 Hz; imaging: $V_s$ = 20 mV, $I_t$ = 150 pA, CO-functionalized tip). (K–M) Constant-height d$I$/d$V$ maps recorded at the sample voltage biases of $V_s$ = +150 mV, $V_s$ = 0 mV, and $V_s$ = –700 mV, respectively ($V_{ac}$ = 10 mV, $f$ = 455 Hz, CO-functionalized tip). (N–P) DFT LDOS maps above a freestanding relaxed finite-segment 5-jsGNR at energies corresponding to the top of the ZMB, the center of the ZMB, and the bottom (top) of the ZMB (VB) calculated using the model for a pure *s*-wave (left) and pure *p*-wave (right) STM tip (experimental d$I$/d$V$ maps acquired with CO-functionalized tips feature contributions of both *s*- and *p*-wave character). (Q) DFT-LDA DOS for a freestanding uncurved 5-jsGNR

(spectrum broadened by 27 meV Gaussian). (R) DFT-LSDA band structure for a freestanding uncurved 5-jsGNR. All STM experiments were performed at $T = 4.5$ K. LDOS maps are sampled at a height of 3 Å above the atomic plane of the GNR.

We again relied on DFT on a freestanding 5-jsGNR to explore the origin of the metallic band observed in STS experiments. To gain insight from the momentum-resolved band structure, we first analyze an uncurved reference structure, constructed by enforcing periodic boundary conditions on the unit cells. Figures 3Q and R show the DOS and electronic band structure obtained from DFT calculations, which converged to a spin-degenerate band structure, i.e., a nonmagnetic metal (Supporting Information Figure S5 shows the robustness of metallic behavior under hole doping). This demonstrates that the fusion which results in the formation of 5-membered rings along the sawtooth edges of 5-jsGNRs restores the spin degeneracy of the ZMB, leading to the emergence of a single dispersive metallic ZMB (with a bandwidth of $\Delta E_{DFT} = 0.7$ eV) crossing the $E_F$. While the top of the ZMB at $\boldsymbol{k} = \Gamma$ gives way to a sizeable gap that separates the half-fill metallic band from the next fully empty band, the bottom of the metallic ZMB band at $\boldsymbol{k} = X$ overlaps with the top of the fully occupied band below it. The DOS thus features two prominent one-dimensional (1D) van Hove singularities straddling the $E_F$ (Figure 3Q). The peak at $E–E_F = -0.25$ eV is significantly larger as it captures both the bottom of the ZMB and the top of the next occupied band below it, a characteristic feature that is reflected in the experimental differential conductance spectra (*feature* 6 in Figure 3J). For comparison with the experimental constant-height d$I$/d$V$ maps, the calculated LDOS maps (Figures 3N–P) are obtained from a relaxed finite segment containing eight unit cells, which displays a curvature closely matching that observed experimentally (Supporting Information Figure S6 provides a detailed analysis of bond lengths and the changes in eigenenergies associated with curvature in 5-jsGNRs).

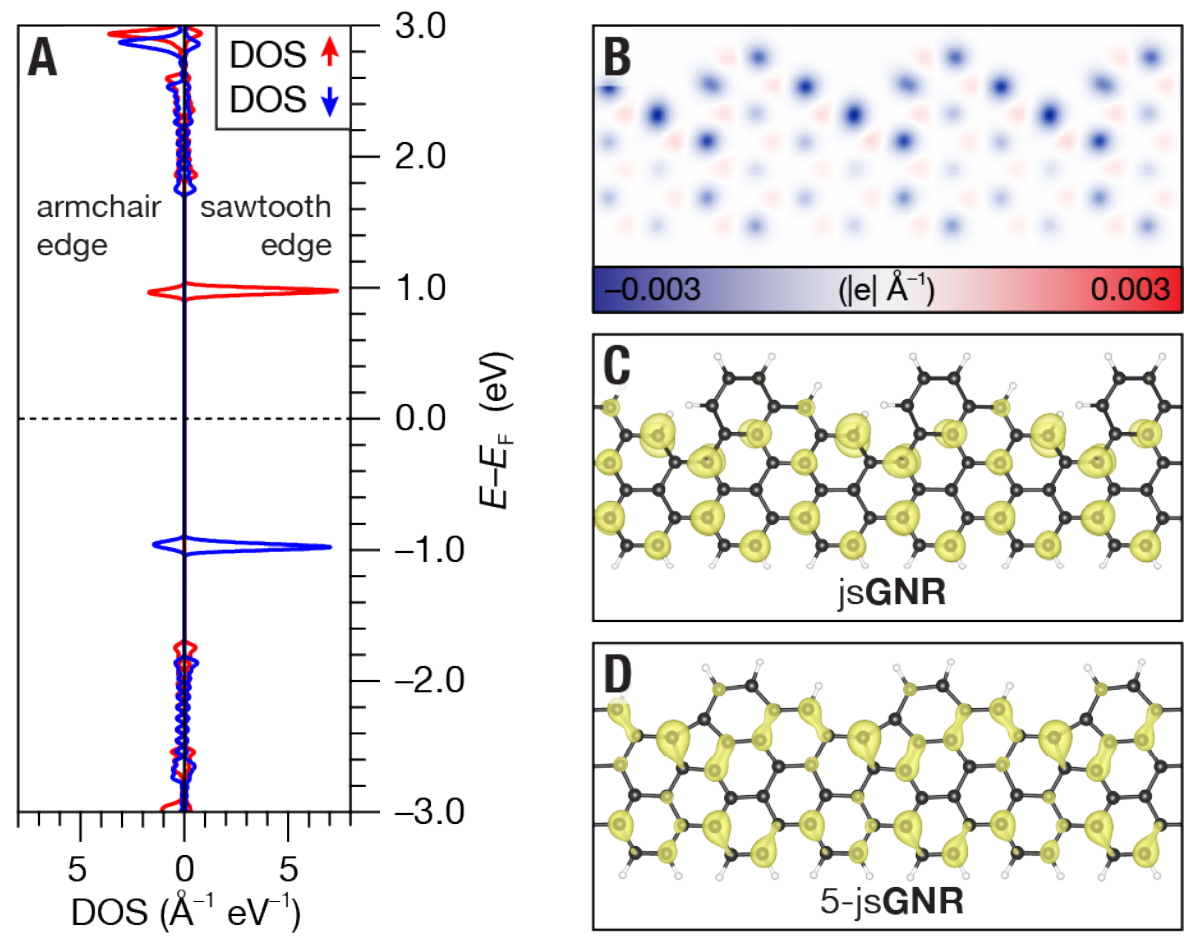

**Figure 4.** Spatial distribution of spin-resolved density of states, net spin densities and band-resolved charge densities in jsGNR and 5-jsGNR. (A) GW-computed PDOS of up (red) and down (blue) spins integrated over the armchair and sawtooth edges of a jsGNR. (B) Spatial distribution of the areal spin density distribution difference of the occupied states ($\rho\uparrow(r)–\rho\downarrow(r)$) in jsGNRs. The areal density is the density integrated in the direction out of the atomic plane. (C) Isosurface representing the DFT-calculated wave function square of jsGNRs at $\boldsymbol{k} = \Gamma$ of the OZMB (surface isovalue set to 5% of the maximum). (D) Isosurface representing the DFT-calculated wave function square of jsGNRs at $\boldsymbol{k} = \Gamma$ of 5-jsGNRs corresponding to the metallic ZMB (surface isovalue set to 5% of the maximum).

***Ab initio* Calculation of the Electronic and Magnetic Structure of jsGNRs and 5-jsGNRs.** *Ab initio* calculations on jsGNRs showed that a ferromagnetic phase of aligned spin density along the sawtooth edge in the ground state is favored over a non-magnetic phase. The calculated spin-polarization energy, 32.7 meV per unit cell, is large and comparable to that of the ferromagnetic ordering calculated along the edges of ZGNRs. Figure 4A shows the spin-resolved partial density of states (PDOS), with up (red) and down (blue) components, integrated over the atoms at the armchair and sawtooth edges of the jsGNR. A map of the net spin density, integrated over all occupied states, reveals that the spin density is primarily localized along the short zigzag segments lining the sawtooth edge (Figure 4B). To gain a more intuitive picture of the interplay between geometric/chemical structure, the effective orbital hopping, and the Stoner instability, we fitted the ZMBs obtained from spin-unpolarized DFT to a 1D TB Hamiltonian. This mapping yields a very small effective nearest-neighbor hopping $t_{1,\mathrm{eff}} = 0.1$ meV, and a larger effective second-nearest-neighbor hopping $t_{2,\mathrm{eff}} = 0.9$ meV. Small contributions from the kinetic energy in jsGNRs are consistent with a localization of the wave function on the majority A sublattice (Figure 4C). The hopping term is dominated by contributions from second-nearest neighbor interactions. In contrast, a TB fit for 5-jsGNR shows dramatically enhanced effective hopping amplitudes between ZMs, with $t_{1,\mathrm{eff}} = 171.5$ meV and $t_{2,\mathrm{eff}} = 25.6$ meV, reflected in the introduction of five-membered rings along the sawtooth edge. The fusion of [4]helicene fragments enhance nearest-neighbor A–A site coupling (linking two A sites by a C–C bond), which delocalizes the ZMB wavefunction to the minority B sites (Figures 4D). In this case, the effective hopping features sizeable contributions from both nearest- and next-nearest-neighbor terms.

In an effort to quantify the electron-electron interaction parameter U, we performed DFT calculations using the constrained Random Phase Approximation (cRPA).[31] The Coulomb repulsion for both the jsGNRs and 5-jsGNRs is found to be $U \sim 200$ meV. Combining the effective hopping terms and the calculated on-site electron-electron repulsion in non-centrosymmetric GNRs, we can now evaluate the Stoner criterion (Eq. 3) for both 1D systems. In the case of jsGNRs the $U$ is much larger than the $t_{\mathrm{eff}}$ ($U \sim 10^2\ t_{\mathrm{eff}}$) and the Stoner instability drives the system towards a magnetically ordered ground state ($U/t_{\mathrm{eff}} >> 2\pi$). The large exchange interaction gives rise to significant spin splitting as reflected in the opening of an experimental electronic band gap observed in STS (Figure 3A). This finding is fully consistent with a ferromagnetic ground state found in spin-polarized DFT-GW calculations (Figure 3I). Performing the same analysis for the system of 5-jsGNRs ($U \sim t_{\mathrm{eff}}$), the larger kinetic energy term overcomes the exchange interaction and the Stoner instability is suppressed ($U/t_{\mathrm{eff}} < 2\pi$), favoring instead a nonmagnetic metallic ground state. This prediction is corroborated by experimental STS (Figure 3J) and DFT-LSDA theory (Figure 3R) which both show a dispersive partially filled metallic band crossing $E_F$.

## CONCLUSIONS

We have demonstrated the rational design, synthesis, and electronic characterization of a new class of non-centrosymmetric GNRs. In jsGNRs, the small effective hopping integral ($t_{\mathrm{eff}} < 1$ meV) between sublattice spin-polarized ZMs leads to an exceptionally high density of states at the $E_F$. The Stoner criterion ($U \times D(E_F) > 1$) is satisfied, triggering a magnetic instability that favors a ferromagnetically ordered insulating ground state ($E_g \sim 1.2$ eV). Sublattice mixing induced by a controlled thermal cyclodehydrogenation of the sawtooth edges (formation of 5-membered rings) increases the effective nearest-neighbor hopping amplitude between ZM orbitals by two orders of magnitude, significantly reducing $D(E_F)$. In 5-jsGNRs, the kinetic energy overcomes the electron-electron repulsion, thereby suppressing the Stoner instability, restoring the spin-degeneracy, and giving rise to a half-filled nonmagnetic metallic ZM band crossing $E_F$. Bond-resolved STM, STS, *ab initio* DFT, cRPA and GW calculations corroborate the irreversible ferromagnetic insulator-to-metal transition. Our results provide a framework for the deterministic design of electronic structure and magnetic ordering in low-dimensional quantum systems using ZM engineering and offer a new platform to design and explore exotic many-body states in graphene nanomaterials.

## METHODS/EXPERIMENTAL

**Precursor Synthesis.** Full details of the synthesis and characterization of molecular precursor **1** are given in the Supporting Information.

**Sample preparation**. Atomically-clean Au(111) substrates were prepared using alternating $Ar^+$ sputtering and annealing cycles under UHV conditions. Sub-monolayer coverages of **1** were prepared via UHV deposition using a Knudsen cell evaporator operating at crucible temperatures of 130 °C for 10 min with the substrate held at 24 °C. Polymerization of **1** was achieved by ramping the substrate temperature by ~3 K $min^{-1}$ to $T$ = 180 °C and holding there for 15 min. Cyclodehydrogenation was induced by ramping the substrate temperature by ~5 K $min^{-1}$ to $T$ = 300 °C and holding there for 15 min.

**STM measurements.** The substrate was characterized at each stage of growth using a commercial Scienta-Omicron low-temperature STM operated at $T$ = 4.5 K with a hand-cut PtIr tip. STS measurements were conducted using a lock-in amplifier at a modulation frequency $f$ = 455 Hz and a modulation voltage $V_{ac}$ = 10 mV. d$I$/d$V$ maps were recorded in constant-height mode with CO-functionalized tips and a modulation voltage $V_{ac}$ = 10 mV. BRSTM images were obtained by mapping the $I$ signal at low-bias (0-5 mV) in constant-height scans using a CO-functionalized tip. STS peak parameters were determined by fitting with Lorentzian line shapes, while the average properties of each feature were determined by ~10 spectra obtained from three GNRs using different STM tips. STM images were processed using the Gwyddion software package.[32]

**Calculations.** DFT calculations on molecular structures were performed using the ORCA software package.[33] Gas-phase geometry optimizations were performed using the M06-2X exchange-correlation functional and the def2-SVP basis set.[34, 35] Single-point energy calculations were performed using the M06-2X functionals with a def2-TZVP and AuxJ basis set[36] at the relaxed geometry for the ground state energy.[37-42] Orbital information was edited using Multiwfn software.[43, 44]

First-principles DFT calculations on GNRs were performed using LSDA functional as implemented in the Quantum ESPRESSO package.[45] Norm-conserving (NC) pseudopotentials were employed[46, 47] with an energy cutoff of 100 Ry and Gaussian smearing of 0.002 Ry. To ensure the accuracy of the results, a vacuum layer of 10 Å was applied along all non-periodic directions. The geometric structures were fully relaxed until the forces on all atoms were smaller than 0.01 eV/Å. Band structure calculations were performed using 40 k-points along the periodic direction.

The intrinsic electronic screening in low-dimensional insulators, such as jsGNRs, is weak. To account for many-body self-energy corrections and to capture the correct quasi-particle band gap we performed ab initio GW calculations[48] on jsGNRs using the BerkeleyGW package.[49] We employed a screened Coulomb interaction cutoff of 15 Ry and a 12 × 1 × 1 q-grid with the nonuniform neck subsampling (NNS) method.[50] A total of 800 bands were included in the summations over empty states for both the dielectric function and the self-energy calculations.

In the case of metallic 5-jsGNRs the highly efficient intra-band screening by itinerant electrons drastically suppresses many-body self-energy corrections.[51, 52] GW corrections for such metallic systems only yield negligible changes to the band dispersion.

cRPA calculations for $U$ on both GNR systems were performed using the Abinit package.[53] Projector augmented-wave (PAW) pseudopotentials[47, 54] were employed with an energy cutoff of 15 Ha and Gaussian smearing of 0.002 Ry. The same crystal structures and atomic configurations as those used in the above DFT calculations were adopted. For the dielectric matrix calculation, a cutoff energy of 5 Ha, an 8 × 1 × 1 q-grid, and a total of 600 bands were used.

## ASSOCIATED CONTENT

### Supporting Information

The Supporting Information is available free of charge on the ACS Publications website.

> Materials and methods, synthesis of molecular precursor 1, STM topographic images of jsGNR, statistical analysis of regioselective polymerization in jsGNRs, d$I$/d$V$ maps of jsGNR and 5-jsGNR, computational models describing the effect of doping and curvature in 5-jsGNR, $^1$H-NMR and $^{13}$C NMR spectra for all compounds.

## AUTHOR INFORMATION

### Corresponding Author

* **Felix R. Fischer** - Department of Chemistry, University of California at Berkeley, Berkeley, California 94720, United States; Materials Sciences Division, Lawrence Berkeley National Laboratory, Berkeley, California 94720, United States; Kavli Energy NanoSciences Institute at the University of California, Berkeley and the Lawrence Berkeley National Laboratory, Berkeley, California 94720, United States; Bakar Institute of Digital Materials for the Planet, Division of Computing, Data Science, and Society, University of California Berkeley, Berkeley, California 94720, United States; https://orcid.org/0000-0003-4723-3111; Email: ffischer@berkeley.edu.

* **Steven G. Louie** - Department of Physics, University of California at Berkeley, Berkeley, California 94720, United States; Materials Sciences Division, Lawrence Berkeley National Laboratory, Berkeley, California 94720, United States; https://orcid.org/0000-0003-0622-0170; Email: sglouie@berkeley.edu.

### Authors

**Aidan P. Delgado** - Department of Chemistry, University of California at Berkeley, Berkeley, California 94720, United States; Kavli Energy NanoSciences Institute at the University of California, Berkeley and the Lawrence Berkeley National Laboratory, Berkeley, California 94720, United States; https://orcid.org/0000-0003-0504-6035.

**Michael C. Daugherty** - Department of Chemistry, University of California at Berkeley, Berkeley, California 94720, United States; https://orcid.org/0000-0002-3348-561X.

**Weichen Tang** - Department of Physics, University of California at Berkeley, Berkeley, California 94720, United States; Materials Sciences Division, Lawrence Berkeley National Laboratory, Berkeley, California 94720, United States; https://orcid.org/0000-0002-1945-5369.

### Author Contributions

‡A.P.D, M.C.D., and W.T. contributed equally to this study.

### Notes

The authors declare no competing financial interest.

## ACKNOWLEDGMENT

This work was primarily funded by the Department of Defense, Office of Naval Research under contract N00014-24-1-2134 (molecular design and synthesis), N00014-19-1-2503 (STM instrumentation), and the US Department of Energy (DOE), Office of Science, Basic Energy Sciences (BES), Materials Sciences and Engineering Division, under contract DE-AC02-05-CH11231 (Nanomachine program KC1203) (on-surface synthesis, DFT calculations) and contract DE-SC0023105 (STM imaging and spectroscopy). This work is also supported by the Center for Computational Study of Excited State Phenomena in Energy Materials (C2SEPEM) funded by the U.S. Department of Energy, Office of Science, Basic Energy Sciences, Materials Sciences and Engineering Division under Contract No. DE-AC02-05CH11231, as part of the Computational Materials Sciences Program (GW calculations and advanced codes). Computational resources were provided by the National Energy Research Scientific Computing Center (NERSC), which is supported by the Office of Science of the U.S. Department of Energy under Contract No. DEAC02-05CH11231. M.C.D. acknowledges a National Defense Science and Engineering Graduate Fellowship. F.R.F. acknowledges generous support by the Heising-Simons Faculty Fellows Program at UC Berkeley. We thank Dr. Hasan Çelik and the UC Berkeley NMR facility in the College of Chemistry (CoC-NMR) for assistance with spectroscopic characterization. Instruments in the CoC-NMR are supported in part by National Institutes of Health (NIH) award no. S10OD024998. We thank Dr. Zhongrui Zhou at the UC Berkeley QB3 Chemistry Mass Spectrometry Facility for assistance with mass spectroscopic characterization.

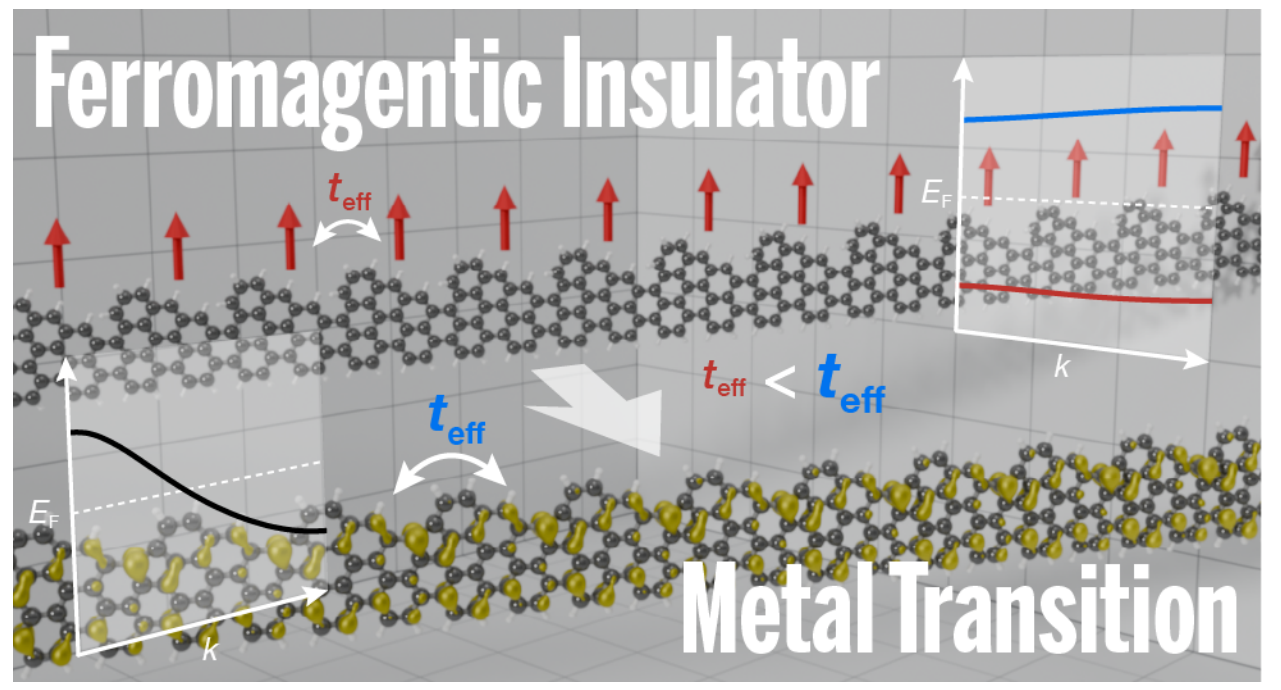